\documentstyle[amsmath,epsf,a4,11pt]{article}
\begin{document}

\date{}
\title{{\bf A relativistic approach to nonlinear peculiar
velocities and the Zeldovich approximation}}
\author{G.F.R. Ellis\thanks{e-mail address: ellis@maths.uct.ac.za}
and C.G. Tsagas\thanks{e-mail address: ctsagas@maths.uct.ac.za}\\
{\small Department of Mathematics and Applied Mathematics,
University of Cape Town}\\
{\small Rondebosch 7701, South Africa}}
\maketitle

\begin{abstract}
We study the peculiar motion of non-relativistic matter in a fully
covariant way. The exact nonlinear equations are derived and then
applied to the case of pressure-free matter, moving relatively to
a quasi-Newtonian Eulerian frame. Our two-frame formalism
facilitates the study of the nonlinear kinematics of the matter,
as the latter decouples from the background expansion and starts
to ``turn around'' and collapse. Applied to second perturbative
order, our equations provide a fully covariant formulation of the
Zeldovich approximation, which by construction addresses the
mildly nonlinear regime of structure formation. Employing a
dynamical system approach, we show that, just like in the
Newtonian case, the relativistic treatment also predicts that
pancakes are the natural end-structures for any generic
overdensity.\\\\ PACS number(s): 04.25.Nx, 98.80.Hw
\end{abstract}

\section{Introduction}
Studies of linear perturbations are crucial for understanding the
way structure formation has progressed in our universe. In the
standard model density fluctuations grow slowly, via gravitational
instability, to form the galaxies, the clusters of galaxies, the
superclusters, the filaments and the voids that we see in the
universe today. Despite the simplicity of this idea, however, our
analytical understanding of the advanced stages of gravitational
collapse is still limited. The reason is that the linear theory is
a good approximation only at the initial stages of the collapse,
when the density contrast of the perturbation is well below unity.
Most of the structures in the universe, however, have density
contrasts well in excess of unity. To understand the evolution of
these objects we need to go beyond the limits of the linear
approximation. The inherent complications of nonlinear collapse,
however, mean that analytical studies are possible only when
certain simplifying assumptions are made. An approximate approach
that extends well into the nonlinear regime, up to the
virialization of bound objects, is the spherical collapse
model~\cite{GG} (see also~\cite{Pee1}). Although the spherical
model became popular because of its simplicity, in reality it
stops short of explaining key features of the observed universe.
Gravitational collapse does not seem to proceed isotropically. All
galaxy surveys show structures with complicated triaxial shapes,
which require a nonspherical analysis if they were to be
explained.

The Zeldovich approximation is not restricted to spherical
symmetry~\cite{Z}. It is a kinematical, Lagrangian approach that
addresses the issue of anisotropic collapse by extrapolating into
the nonlinear regime a well known linear result (see~\cite{SZ} for
a review and also~\cite{P} for related discussion). More
specifically, the Zeldovich ansatz assumes that the
acceleration-free and irrotational linear peculiar motion of
pressure-free matter also holds during the nonlinear collapse. The
consequences of this hypothesis are quite dramatic. What Zeldovich
showed was that any generic (i.e.~non-spherical) overdensity will
undergo a phase of anisotropic, effectively one-dimensional,
collapse leading to the formation of two-dimensional flattened
structures that are widely known as ``pancakes''. Over the years,
the Zeldovich approximation has provided a great deal of insight
into the initial nonlinear evolution of density fluctuations. It
has also inspired a number of developments concerning the behavior
of the velocity field in irrotational flows (see~\cite{NDBB} for a
representative list).\footnote{The limits of the Zeldovich
approximation, both linear and nonlinear, have been investigated
in~\cite{Bu1}. Some authors have suggested modifications of the
standard Lagrangian approach to increase its accuracy, by adding
vorticity~\cite{BS}, or viscosity~\cite{KPS}, while others have
treated either the velocity potential as a constant~\cite{MLMS},
or the gravitational potential itself~\cite{BP}.} Most of the
available studies, however, are Newtonian and, although there are
general relativistic approaches in the literature (see \cite{K}),
a fully covariant treatment is still missing. Moreover, Newtonian
and relativistic collapse seem to move further apart in the
presence of anisotropy (e.g.~see~\cite{ZN}). In the Newtonian
approach pancakes are the natural attractors, at least in the
Zeldovich approximation; a result that seems well confirmed by
N-body simulations~\cite{SMMPT}. The relativistic ``silent
universe'' models, on the other hand, point towards spindle-like
singularities~\cite{BJ, BMP1}.\footnote{Silent universes are
inhomogeneous spacetimes, with irrotational pressure-free matter
and zero magnetic Weyl tensor. By construction, they cannot
support any communication through sound or gravitational wave
signals, which explains the term silent. This class of models
contains the well known Szekeres solution as a special case. For
an introduction and further discussion on silent universes we
refer the reader to~\cite{MPS}.} In other words, the final fate of
a collapsing overdensity remains an open issue. In the present
article we attempt to address this question by looking at the
dynamics of nonlinear collapse in a perturbed Einstein-de Sitter
universe.

One issue that emerges naturally, especially when attempting a
relativistic analysis of peculiar motions, is that the associated
velocities should be defined relative to a preferred frame of
reference. This frame is necessarily non-comoving with the matter,
since by definition there are no peculiar velocities relative to
the matter frame, which is the truly Lagrangian frame. It should
be emphasized that in the standard literature the term
``comoving'' refers usually (and sometimes misleadingly) to a
fictitious background 4-velocity, rather than to the actual
velocity of the matter. In the covariant approach one seeks to
define the aforementioned preferred frame in a physical
way.\footnote{An extensive presentation of the covariant approach
to cosmology can be found in~\cite{E1}. For an updated review the
reader is referred to~\cite{EvE}.} Motivated
by~\cite{Pee}-\cite{EvEM}, we chose a 4-velocity field that is
both irrotational and shear-free, thus defining a quasi-Newtonian
frame that closely corresponds to Bardeen's quasi-Newtonian
gauge~\cite{B}. Our aim is to set up the general framework for the
fully covariant and fully nonlinear treatment of non-relativistic
peculiar motions. We then exploit the natural transparency of the
covariant formalism to provide a fully covariant version of the
Zeldovich approximation. As is well known, the latter addresses
the early stages of nonlinear collapse. We call this period the
``mildly nonlinear regime''. Given that we deal with the
post-recombination era, we assume an Einstein-de Sitter background
and consider the second order evolution of the perturbed
variables. In the frame of the pressure-free matter, we find that
the equations describing the collapse of the non-relativistic
component to second perturbative order do not include any tidal
effects. They reduce to a set of ordinary second order
differential equations, which can be rescaled into a planar
dynamical system with one-dimensional pancakes as its natural
attractors.

\section{The non-linear peculiar kinematics}
\subsection{The peculiar velocity field}
Velocity perturbations and their implications for gravitational collapse are
fundamental in any structure formation scenario. In the post-recombination
era and on scales well below the Hubble length, the Newtonian theory can
adequately address the question of gravitational instability. As we are
successively probing significant fractions of the Hubble radius, however,
the need for a relativistic treatment grows. In what follows we will lay
down the formalism for the fully covariant treatment of peculiar velocities
in cosmology.

To begin with, consider a family of timelike worldlines tangent to
the reference 4-velocity vector field $u_a$, normalized so that
$u_au^a=-1$, and assume the presence of matter moving with
4-velocity $\tilde{u}_a$ so that
\begin{equation}
\tilde{u}_a=\gamma(u_a+ v_a)\,.  \label{utilde}
\end{equation}
Here $v_a$ is the peculiar velocity of the matter with respect to
$u_a$, $ \gamma=(1-v^2)^{-1/2}$ is the Lorentz-boost factor
between the two frames and $v^2=v_av^a$. Note that $v_a$ vanishes
in the background, because in a Robertson-Walker model both $u_a$
and $\tilde{u}_a$ would be chosen parallel to the canonical time
direction, which makes it a first order gauge-invariant quantity.
Also, for non-relativistic peculiar motions, like those we will be
dealing with, $v^2\ll1$. This means that $\gamma\simeq1$ , which
in turn ensures that $\tilde{u}_a\tilde{u}^a\simeq-1$. Apart from
this restriction, the frames $u_a$ and $\tilde{u}_a$ are
completely general. Later, however, they will be respectively
identified with the quasi-Newtonian observers and with the
(pressure-free) matter (see Sec.~3).

The 4-velocity field $\tilde{u}_a$ of the matter can be used to
define an additional time direction and an associated projection
tensor given by
\begin{equation}
\tilde{h}_{ab}=g_{ab}+\tilde{u}_a\tilde{u}_b  \label{tilde-hab}
\end{equation}
where $g_{ab}$ is the spacetime metric. Note that $v_a$ is not
orthogonal to $\tilde{u}_a$. Indeed, even for $\gamma=1$,
Eqs.~(\ref{utilde}) and (\ref{tilde-hab}) ensure that
$\tilde{u}_av^a=v^2\neq0$ and that $\tilde{h}_{ab}v^b=v_a+v^2
\tilde{u}_a\neq v_a$ respectively.

\subsection{Decomposition of the peculiar velocity gradients}
Following the standard covariant splitting of the 4-velocity
gradients into the irreducible kinematical quantities of the
motion, we will also decompose the gradients of the peculiar
velocity field. We take $\tilde{u}_a$ as our time direction
because this leads to simpler equations later on (in effect we are
choosing a non hypersurface orthogonal threading of the space-time
manifold). We then arrive at the expression
\begin{equation}
\nabla_bv_a=\tilde{{\rm D}}_bv_a- \dot{v}_a\tilde{u}_b+
\tilde{u}_a(\tilde{\sigma}_{bc}-\tilde{\omega}_{bc}
+{\textstyle{\frac{1}{3}}}\tilde{\Theta}\tilde{h}_{bc})v^c-
\tilde{u}_a\nabla_bv^2- (v^2)^{.}\tilde{u}_a\tilde{u}_b\,,
\label{nabla-v}
\end{equation}
where $\dot{v}_a=\tilde{u}^b\nabla_bv_a$ and $\tilde{{\rm
D}}_a=\tilde{h}_a{}^b\nabla_b$ is the covariant derivative
operator orthogonal to $\tilde{u}_a$. It should be emphasized that
$\tilde{\Theta}$, $\tilde{\sigma}_{ab}$ and $\tilde{\omega}_{ab}$
are the expansion scalar, the shear and the vorticity associated
with $\tilde{u}_a$. Also, in deriving Eq.~(\ref{nabla-v}) we have
used the relation $\tilde{u}_av^a=v^2$, which holds for
$\gamma=1$.

Similarly, the projected gradient $\tilde{{\rm D}}_bv_a$ of the
peculiar velocity is decomposed into the irreducible quantities of
the motion as follows
\begin{equation}
\tilde{{\rm D}}_bv_a=
{\textstyle{\frac{1}{3}}}\hat{\Theta}\tilde{h}_{ab}+
\hat{\sigma}_{ab}+ \hat{\omega}_{ab}\,,  \label{Dbva-split}
\end{equation}
where\footnote{Round brackets indicate symmetrization and square
ones antisymmetrization. Angled brackets, on the other hand,
indicate the projected, symmetric and trace-free part of tensors
and also the projected component of vectors.}
\begin{equation}
\hat{\Theta}\equiv\tilde{{\rm D}}_av^a\,, \hspace{1cm}
\hat{\sigma}_{ab}\equiv\tilde{{\rm D}}_{\langle b}v_{a\rangle}
\hspace{5mm}{\rm and} \hspace{5mm}
\hat{\omega}_{ab}\equiv\tilde{{\rm D}}_{[b}v_{a]}\,.
\label{hatvars}
\end{equation}
The above respectively represent what one might call the expansion
(or contraction) scalar, the shear tensor and the vorticity tensor
associated with the peculiar motion. Clearly, when
$\hat{\Theta}>0$ we are dealing with an expanding fluid element,
whereas $\hat{\Theta}<0$ indicates collapse.

\subsection{The peculiar motion}
The motion of the matter is monitored through a set of two fully
nonlinear propagation equations, with time derivatives taken along
the matter flow lines. The first determines the evolution of the
peculiar velocity vector
\begin{equation}
\dot{v}_{\langle a\rangle}=\tilde{A}_a- A_{\langle a\rangle}-
{\textstyle{\frac{1}{3}}}(\tilde{\Theta}-\hat{\Theta})v_{\langle
a\rangle}- (\tilde{\sigma}_{ab}-\hat{\sigma}_{ab})v^b-
(\tilde{\omega}_{ab}- \hat{\omega}_{ab})v^b\,,  \label{va-prop}
\end{equation}
and the second governs the behavior of its projected gradients
\begin{eqnarray}
\tilde{h}_a{}^c\tilde{h}_b{}^d\left(\tilde{{\rm
D}}_dv_c\right)^{.}&=&\tilde{{\rm D}}_b\dot{v}_a-
\left(\tilde{\sigma}^c{}_b+\tilde{\omega}^c{}_b+
{\textstyle{\frac{1}{3}}}\tilde{\Theta}\tilde{h}^c{}_b\right)
\left(\hat{\sigma}_{ca}+\hat{\omega}_{ca}
+{\textstyle{\frac{1}{3}}}\hat{\Theta}\tilde{h}_{ca}\right)+
\dot{v}_{\langle a\rangle}\tilde{A}_b\nonumber\\&{}&
-\tilde{A}_a\left(\tilde{\sigma}_{cb}+\tilde{\omega}_{cb}
+{\textstyle{\frac{1}{3}}}\tilde{\Theta}\tilde{h}_{cb}\right)v^c+
\tilde{A}_a{\rm D}_bv^2-
\tilde{h}_a{}^c\tilde{h}_b{}^dR_{ecsd}v^e\tilde{u}^s\,.
\label{Dbva-prop}
\end{eqnarray}
The former of the two equations can also be seen as the
transformation law of the 4-acceleration between the ``tilded''
and the ``non-tilded'' frame, with
$\tilde{A}_a=\tilde{u}^b\nabla_b\tilde{u}_a$ and
$A_a=u^b\nabla_bu_a$. The latter comes after applying the Ricci
identity to $v_a$, projecting orthogonal to $\tilde{u}_a$ and then
employing decompositions (\ref{nabla-v}) and (\ref{Dbva-split}).

The trace, the symmetric trace-free part and the antisymmetric
component of Eq.~(\ref{Dbva-prop}) provide the nonlinear evolution
equations for the irreducible kinematical variables of the
peculiar motion. In particular we have
\begin{eqnarray}
\dot{\hat{\Theta}}&=&
-{\textstyle{\frac{1}{3}}}\tilde{\Theta}\hat{\Theta}-
\tilde{\sigma}_{ab}\hat{\sigma}^{ab}+
\tilde{\omega}_{ab}\hat{\omega}^{ab}+ \tilde{{\rm D}}_a\dot{v}^a+
\tilde{A}^a\dot{v}_a- R_{ab}v^a\tilde{u}^b\nonumber\\&{}&
-\left(\tilde{\sigma}_{ab}-\tilde{\omega}_{ab}+
{\textstyle{\frac{1}{3}}}\tilde{\Theta}\tilde{h}_{ab}\right)
\tilde{A}^av^b+ \tilde{A}^a\tilde{{\rm D}}_av^2  \label{p-Ray}
\end{eqnarray}
for the peculiar expansion (or contraction),
\begin{eqnarray}
\tilde{h}_{\langle a}{}^c\tilde{h}_{b\rangle}{}^d
\dot{\hat{\sigma}}_{cd}&=& -{\textstyle{\frac{1}{3}}}
(\tilde{\Theta}\hat{\sigma}_{ab} +\hat{\Theta}
\tilde{\sigma}_{ab})- \tilde{\sigma}_{c\langle
a}\hat{\sigma}_{b\rangle}{}^c- \tilde{\omega}_{c\langle
a}\hat{\omega}_{b\rangle}{}^c+ \tilde{{\rm D}}_{\langle
a}\dot{v}_{b\rangle}- \tilde{A}_{\langle
a}h_{b\rangle}{}^c\dot{v}_c- \tilde{h}_{\langle
a}{}^c\tilde{h}_{b\rangle}{}^dR_{ecsd}v^e \tilde{u}^s
\nonumber\\&{}& -\tilde{\sigma}_{c\langle
a}\hat{\omega}_{b\rangle}{}^c- \tilde{\omega} _{c\langle
a}\hat{\sigma}_{b\rangle}{}^c- \tilde{A}_{\langle
a}\left(\tilde{\sigma}_{b\rangle c}- \tilde{\omega}_{b\rangle
c}+{\textstyle{\frac{1}{3}}}\tilde{\Theta}\tilde{h}_{b\rangle
c}\right)v^c+ \tilde{A}_{\langle a}\tilde{{\rm D}}_{b\rangle}v^2
\label{p-shear-prop}
\end{eqnarray}
for the peculiar shear and
\begin{eqnarray}
\tilde{h}_{[a}{}^c\tilde{h}_{b]}{}^d\dot{\hat{\omega}}_{cd}&=&
-{\textstyle{\frac{1}{3}}}(\tilde{\Theta}\hat{\omega}_{ab}
+\hat{\Theta}\tilde{\omega}_{ab})+
\tilde{\sigma}_{c[a}\hat{\sigma}_{b]}{}^c+
\tilde{\omega}_{c[a}\hat{\omega}_{b]}{}^c- \tilde{{\rm
D}}_{[a}\dot{v}_{b]}- \tilde{A}_{[a}h_{b]}{}^c\dot{v}_c-
\tilde{h}_{[a}{}^c\tilde{h}_{b]}{}^dR_{ecsd}v^e\tilde{u}^s
\nonumber\\&{}& +\tilde{\sigma}_{c[a}\hat{\omega}_{b]}{}^c+
\tilde{\omega}_{c[a}\hat{\sigma}_{b]}{}^c-
\tilde{A}_{[a}\left(\tilde{\sigma}_{b]c}-\tilde{\omega}_{b]c}
+{\textstyle{\frac{1}{3}}}\tilde{\Theta}\tilde{h}_{b]c}\right)v^c+
\tilde{A}_{[a}\tilde{{\rm D}}_{b]}v^2\,.  \label{p-vort-prop}
\end{eqnarray}
for the peculiar vorticity.

\subsection{The spacetime curvature effects}
The curvature terms in the right hand side of
Eqs.~(\ref{p-Ray})-(\ref {p-vort-prop}) convey the gravitational
effects on the peculiar kinematics. To evaluate their contribution
we consider the standard decomposition of the Riemann tensor
\begin{equation}
R_{abcd}=C_{abcd}+
{\textstyle{\frac{1}{2}}}(g_{ac}R_{bd}+g_{bd}R_{ac}
-g_{bc}R_{ad}-g_{ad}R_{bc})-
{\textstyle{\frac{1}{6}}}R(g_{ac}g_{bd}-g_{ad}g_{bc})\,,
\label{Riemann}
\end{equation}
where $R_{ab}=R^c{}_{acb}$ and $R=R^a{}_a$ are the Ricci tensor
and the Ricci scalar respectively, while $C_{abcd}$ is the Weyl
tensor. The latter describes the long range gravitational field,
namely tidal forces and gravity waves. Relative to $ \tilde{u}_a$
the Weyl tensor decomposes further as follows~\cite{M1}
\begin{eqnarray}
C_{abcd}&=&\tilde{u}_a\tilde{u}_c\tilde{E}_{bd}-
\tilde{u}_a\tilde{u}_d \tilde{E}_{bc}-
\tilde{u}_b\tilde{u}_c\tilde{E}_{ad}+ \tilde{u}_b\tilde{u}_d
\tilde{E}_{ac}-
\tilde{\varepsilon}_{abe}\tilde{\varepsilon}_{cdf}\tilde{E}^{ef}
\nonumber\\&{}&
+\tilde{\varepsilon}_{abe}\tilde{u}_c\tilde{H}^e{}_d-
\tilde{\varepsilon}_{abe}\tilde{u}_d\tilde{H}^e{}_c+
\tilde{\varepsilon}_{cdf}\tilde{u}_a \tilde{H}_b{}^f-
\tilde{\varepsilon}_{cdf}\tilde{u}_b\tilde{H}_a{}^f\,,
\label{Weyl}
\end{eqnarray}
where $\tilde{E}_{ab}$ and $\tilde{H}_{ab}$ are the associated
electric and magnetic Weyl components. Note that
$\tilde{\varepsilon}_{abc}=\eta_{abcd}\tilde{u}^d$ is the totally
antisymmetric 3-dimensional permutation tensor orthogonal to
$\tilde{u}_a$, with $\eta_{abcd}$ being the spacetime alternating
tensor. Introducing the above into expression (\ref {Riemann}) we
obtain
\begin{equation}
R_{ab}v^a\tilde{u}^b=-\tilde{q}_av^a-
\left[{\textstyle{\frac{1}{2}}}(\tilde{\mu}+3\tilde{p})-\Lambda\right]v^2\,,
\label{SC1}
\end{equation}
\begin{equation}
\tilde{h}_{\langle a}{}^c\tilde{h}_{b\rangle}{}^d
R_{ecsd}v^e\tilde{u}^s= -v^2\tilde{E}_{ab}+
\tilde{\varepsilon}_{cd\langle a}v^c\tilde{H}^d{}_{b\rangle}+
{\textstyle{\frac{1}{2}}}v^2\tilde{\pi}_{ab}+
{\textstyle{\frac{1}{2}}}\tilde{q}_{\langle
a}\tilde{h}_{b\rangle}{}^cv_c\,, \label{SC2}
\end{equation}
and
\begin{equation}
\tilde{h}_{[a}{}^c\tilde{h}_{b]}{}^dR_{ecsd}v^e\tilde{u}^s=
\tilde{\varepsilon}_{cd[a}v^c\tilde{H}^d{}_{b]}+
{\textstyle{\frac{1}{2}}}\tilde{q}_{[a}\tilde{h}_{b]}{}^cv_c\,,
\label{SC3}
\end{equation}
where $\tilde{\mu}$, $\tilde{p}$, $\tilde{q}_a$ and
$\tilde{\pi}_{ab}$ are respectively the energy density, the
isotropic pressure, the heat flux and the anisotropic stresses of
any matter fields that may be present as measured in the
$\tilde{u}_a$ frame.

\subsection{The fully nonlinear equations}
Substituting results (\ref{SC1})-(\ref{SC3}) into
Eqs.~(\ref{p-Ray})-(\ref {p-vort-prop}) we obtain the following
fully nonlinear expressions for the irreducible kinematical
variables
\begin{eqnarray}
\dot{\hat{\Theta}}&=&-
{\textstyle{\frac{1}{3}}}\tilde{\Theta}\hat{\Theta}-
\tilde{\sigma}_{ab}\hat{\sigma}^{ab}+
\tilde{\omega}_{ab}\hat{\omega}^{ab}+ \tilde{{\rm D}}_a\dot{v}^a+
\tilde{A}^a\dot{v}_a+ \tilde{q}_av^a+
\left[{\textstyle{\frac{1}{2}}}(\tilde{\mu}+3\tilde{p})-\Lambda\right]v^2
\nonumber\\&{}& -\left(\tilde{\sigma}_{ab}-\tilde{\omega}_{ab}
+{\textstyle{\frac{1}{3}}}\tilde{\Theta}\tilde{h}_{ab}\right)\tilde{A}^av^b+
\tilde{A}^a\tilde{{\rm D}}_av^2\,,  \label{p-Ray2}
\end{eqnarray}
\begin{eqnarray}
\tilde{h}_{\langle a}{}^c\tilde{h}_{b\rangle}{}^d
\dot{\hat{\sigma}}_{cd}&=&-
{\textstyle{\frac{1}{3}}}(\tilde{\Theta}\hat{\sigma}_{ab}
+\hat{\Theta}\tilde{\sigma}_{ab})- \tilde{\sigma}_{c\langle
a}\hat{\sigma}_{b\rangle}{}^c- \tilde{\omega}_{c\langle
a}\hat{\omega}_{b\rangle}{}^c+ \tilde{{\rm D}}_{\langle
a}\dot{v}_{b\rangle}- \tilde{A}_{\langle
a}h_{b\rangle}{}^c\dot{v}_c+ v^2\tilde{E}_{ab}-
\tilde{\varepsilon}_{cd\langle a}v^c\tilde{H}^d{}_{b\rangle}
\nonumber\\&{}& -{\textstyle{\frac{1}{2}}}v^2\tilde{\pi}_{ab}-
{\textstyle{\frac{1}{2}}}\tilde{q}_{\langle
a}\tilde{h}_{b\rangle}{}^cv_c- \tilde{\sigma}_{c\langle a}
\hat{\omega}_{b\rangle}{}^c- \tilde{\omega}_{c\langle
a}\hat{\sigma}_{b\rangle}{}^c- \tilde{A}_{\langle
a}\left(\tilde{\sigma}_{b\rangle c}- \tilde{\omega}_{b\rangle c}
+{\textstyle{\frac{1}{3}}}\tilde{\Theta}\tilde{h}_{b\rangle
c}\right)v^c \nonumber\\&{}& +\tilde{A}_{\langle a}\tilde{{\rm
D}}_{b\rangle}v^2\,. \label{p-shear-prop2}
\end{eqnarray}
and
\begin{eqnarray}
\tilde{h}_{[a}{}^c\tilde{h}_{b]}{}^d\dot{\hat{\omega}}_{cd}&=&
-{\textstyle{\frac{1}{3}}}(\tilde{\Theta}\hat{\omega}_{ab}
+\hat{\Theta}\tilde{\omega}_{ab})+
\tilde{\sigma}_{c[a}\hat{\sigma}_{b]}{}^c+
\tilde{\omega}_{c[a}\hat{\omega}_{b]}{}^c- \tilde{{\rm
D}}_{[a}\dot{v}_{b]}- \tilde{A}_{[a}h_{b]}{}^c\dot{v}_c-
\tilde{\varepsilon}_{cd[a}v^c\tilde{H}^d{}_{b]}-
{\textstyle{\frac{1}{2}}}\tilde{q}_{[a}\tilde{h}_{b]}{}^cv_c
\nonumber\\&{}& +\tilde{\sigma}_{c[a}\hat{\omega}_{b]}{}^c+
\tilde{\omega}_{c[a}\hat{\sigma}_{b]}{}^c-
\tilde{A}_{[a}\left(\tilde{\sigma}_{b]c}-\tilde{\omega}_{b]c}
+{\textstyle{\frac{1}{3}}}\tilde{\Theta}\tilde{h}_{b]c}\right)v^c+
\tilde{A}_{[a}\tilde{{\rm D}}_{b]}v^2\,.  \label{p-vort-prop2}
\end{eqnarray}
The above, together with the propagation equation of $v_a$ itself, that is
\begin{equation}
\dot{v}_{\langle a\rangle}=\tilde{A}_a- A_{\langle a\rangle}-
{\textstyle{\frac{1}{3}}}(\tilde{\Theta}-\hat{\Theta})v_{\langle
a\rangle}- (\tilde{\sigma}_{ab}- \hat{\sigma}_{ab})v^b-
(\tilde{\omega}_{ab}- \hat{\omega}_{ab})v^b\,,  \label{va-prop2}
\end{equation}
determine the peculiar motion completely. Finally, we remind the
reader that to this point we have not explicitly identified $u_a$
and $\tilde{u}_a$ with any physical frame. In other words, the
generality of formulae (\ref{p-Ray2})-(\ref{va-prop2}) is only
restricted by the fact that the relative motion of the
aforementioned observers is non-relativistic (i.e.~$\gamma=1$).

\section{Application to pressure-free matter}
\subsection{The quasi-Newtonian observers}
When it comes to studies of relativistic cosmological
perturbations, Bardeen's gauge-invariant theory is the most
popular approach~\cite{B}. Various gauge choices are possible in
this formalism. The longitudinal gauge for example, which assumes
constant-time slices with zero-shear normals, sets up a frame of
reference that emulates the Eulerian observers of the Newtonian
theory, thus motivating the term ``quasi-Newtonian'' gauge. Since
this frame is non-comoving, one can use it to study the relative
velocity effects. The covariant alternative mimics the fixing
conditions of Bardeen's quasi-Newtonian gauge by introducing an
1+3 threading of the spacetime via a 4-velocity field that is both
irrotational and shear-free~\cite{vEE,EvEM}. Thus, covariantly,
quasi-Newtonian cosmologies are almost-FRW universes equipped with
a family of observers, moving non-relativistically relative to the
comoving observers, whose motion is also non-rotating and
shear-free. In addition to, such cosmological models do not
support gravitational waves, which further justifies the term
quasi-Newtonian~\cite{vEE}.

In the current section, the quasi-Newtonian observers will be
identified with the 4-velocity vector field $u_a$ defined in
Sec.~2. As a result, the shear and the vorticity tensors
associated with $u_a$ will vanish
(i.e.~$\sigma_{ab}=0=\omega_{ab}$). The rest of the kinematic,
dynamic and gravito-electromagnetic quantities depends on the
choice of the matter component and on the transformation laws
between the $u_a$ and the $\tilde{u} _a$ frames. For the full list
of the exact nonlinear relations the reader is referred
to~\cite{M2,MGE}.

\subsection{The nonlinear peculiar motion of the matter}
So far, $v_a$ is an arbitrary peculiar velocity field which has
not been associated with any particular matter source. To this
point, Eqs.~(\ref {p-Ray2})-(\ref{va-prop2}) are completely
general and $\gamma=1$ is the only restriction imposed. Next, we
will align $\tilde{u}_a$ with the motion of a non-relativistic
matter component. This choice allow us to address the question of
velocity perturbations in the post-recombination era, when the
universe is dominated by pressure-free dust, while it considerably
simplifies Eqs.~(\ref{p-Ray2})-(\ref{va-prop2}). In particular,
$\tilde{q} _a=0$ in the matter frame by definition and
$\tilde{p}=\tilde{\pi}_{ab}= \tilde{A}_a=0$ for dust. Therefore,
the nonlinear peculiar motion of the pressureless component is
governed by
\begin{equation}
\dot{v}_{\langle a\rangle}= -A_{\langle a\rangle}
-{\textstyle{\frac{1}{3}}}(\tilde{\Theta}-\hat{\Theta})v_{\langle
a\rangle}- (\tilde{\sigma}_{ab}- \hat{\sigma}_{ab})v^b-
(\tilde{\omega}_{ab}- \hat{\omega}_{ab})v^b\,, \label{va-prop3}
\end{equation}
\begin{equation}
\dot{\hat{\Theta}}=
-{\textstyle{\frac{1}{3}}}\tilde{\Theta}\hat{\Theta}+
\tilde{\sigma}_{ab}\hat{\sigma}^{ab}+
\tilde{\omega}_{ab}\hat{\omega}^{ab}+ \tilde{{\rm D}}_a\dot{v}^a+
\left({\textstyle{\frac{1}{2}}}\tilde{\mu} -\Lambda\right)v^2\,,
\label{p-Ray3}
\end{equation}
\begin{eqnarray}
\dot{\hat{\sigma}}_{ab}&=&
-{\textstyle{\frac{1}{3}}}(\tilde{\Theta}\hat{\sigma}_{ab}
+\hat{\Theta}\tilde{\sigma}_{ab})- \tilde{\sigma}_{c\langle
a}\hat{\sigma}_{b\rangle}{}^c- \tilde{\omega}_{c\langle
a}\hat{\omega}_{b\rangle}{}^c+ \tilde{{\rm D}}_{\langle
a}\dot{v}_{b\rangle}+ v^2\tilde{E} _{ab}-
\tilde{\varepsilon}_{cd\langle a}v^c\tilde{H}^d{}_{b\rangle}
\nonumber\\&{}& -\tilde{\sigma}_{c\langle
a}\hat{\omega}_{b\rangle}{}^c- \tilde{\omega}_{c\langle
a}\hat{\sigma}_{b\rangle}{}^c\,,  \label{p-shear-prop3}
\end{eqnarray}
and
\begin{eqnarray}
\underline{}\dot{\hat{\omega}}_{ab}&=&
-{\textstyle{\frac{1}{3}}}(\tilde{\Theta}\hat{\omega}_{ab}
+\hat{\Theta}\tilde{\omega}_{ab})+
\tilde{\sigma}_{c[a}\hat{\sigma}_{b]}{}^c+
\tilde{\omega}_{c[a}\hat{ \omega}_{b]}{}^c- \tilde{{\rm
D}}_{[a}\dot{v}_{b]}-
\tilde{\varepsilon}_{cd[a}v^c\tilde{H}^d{}_{b]} \nonumber\\&{}&
+\tilde{\sigma}_{c[a}\hat{\omega}_{b]}{}^c+
\tilde{\omega}_{c[a}\hat{\sigma}_{b]}{}^c\,. \label{p-vort-prop3}
\end{eqnarray}

Note the effect of the long range gravitational field on the
anisotropy of the collapse, which propagates through the electric
and magnetic Weyl terms in Eq.~(\ref{p-shear-prop3}). Their
coupling to the peculiar velocity of the matter means that the
tidal effects are of higher perturbative order than the rest. This
will prove crucial when Eqs.~(\ref{va-prop3})-(\ref{p-vort-prop3})
are applied to the mildly nonlinear regime of gravitational
collapse (see Sec.~4.1). Also note how the tidal field acts as a
source of peculiar vorticity through the magnetic Weyl component
in Eq.~(\ref{p-vort-prop3}).

\subsection{The linear peculiar motion of the matter}
For quasi-Newtonian observers $\sigma_{ab}=0=\omega_{ab}$ and the
linear transformation laws between the quasi-Newtonian and the
matter frames guarantee that~\cite{M2,MGE}
\begin{equation}
\tilde{\sigma}_{ab}=\hat{\sigma}_{ab}\,,\hspace{1cm}
\tilde{\omega}_{ab}= \hat{\omega}_{ab} \hspace{5mm} {\rm and}
\hspace{5mm} \tilde{\Theta}=\Theta+ \hat{\Theta}\,.
\label{lin-kin}
\end{equation}
Also to first order, the matter density and the
gravito-electromagnetic quantities are related by
\begin{equation}
\tilde{\mu}=\mu\,,\hspace{1cm} \tilde{E}_{ab}=E_{ab} \hspace{5mm}
{\rm and} \hspace{5mm} \tilde{H}_{ab}=0\,.  \label{lin-grav}
\end{equation}
Finally, the 4-acceleration $A_a$ of the quasi-Newtonian observers
takes the form~\cite{EvEM}
\begin{equation}
A_a=\tilde{{\rm D}}_a\phi\,,  \label{Aa}
\end{equation}
where $\phi$ is the peculiar gravitational potential. The latter
has the trivial gauge freedom to add, or subtract, an arbitrary
homogeneous function. Note that $\tilde{{\rm D}}_a\phi={\rm
D}_a\phi$ to first order, given that $\tilde{h}_{ab}=h_{ab}$ to
zero order. Employing the above results Eq.~(\ref{va-prop3})
reduces to the following expression
\begin{equation}
\dot{v}_a+ {\textstyle{\frac{1}{3}}}\Theta v_a=-\tilde{{\rm D}}_a\phi\,,
\label{lin-va-prop}
\end{equation}
which monitors the linear evolution of the peculiar velocity
between the quasi-Newtonian and the matter frames relative to the
latter. Recall that an overdot indicates covariant differentiation
along $\tilde{u}_a$ (i.e.~$\dot{v }_a=\tilde{u}^b\nabla_bv_a$).

For an Einstein-de Sitter background and relative to the
quasi-Newtonian observers, the linear peculiar velocity has a
single growing mode given by $
v^a=v_a^0a^{1/2}$~\cite{EvEM}.\footnote{ The linear evolution law
$v_a\propto a^{1/2}$ obtained in~\cite{EvEM} assumes that the
peculiar velocity is aligned with the peculiar acceleration,
namely that $v_a\propto A_a$. This assumption, which is also part
of the Zeldovich approximation, ignores contributions from any
component of $v_a$ that happens to be orthogonal to $A_a$. As has
been pointed out in~\cite{EvEM}, this is not unreasonable since:
(i) if the orthogonal component is zero at some initial time, it
remains so; (ii) if there is an initial orthogonal component, then
the expansion and the matter aggregation will serve to decrease
it.} Here, $v_a^0$ is the peculiar velocity at some initial time
and the scale factor has been normalized so that $a_0=1$. Now, to
linear order, it is straightforward to employ Eq.~(\ref {utilde})
and show that $\tilde{u}^b\nabla_bv_a=u^b\nabla_bv_a$, which
immediately implies that $v_a$ has the same linear evolution
(i.e.~$ v_a\propto a^{1/2}$) in both the quasi-Newtonian and the
matter frames. On these grounds, it helps to define the rescaled
peculiar velocity field
\begin{equation}
V_a=a^{-1/2}v_a\,.  \label{Va}
\end{equation}
The advantage of $V_a$ is that, to leading order, it is constant
both in the quasi-Newtonian and in the matter frame by
construction. Thus, $\dot{V}_a=0$ to linear order. An immediate
consequence is that the rescaled peculiar velocity derives from a
potential. Indeed, by substituting (\ref{Va}) into
Eq.~(\ref{lin-va-prop}) we arrive at the linear expression
\begin{equation}
\dot{V}_a+ {\textstyle{\frac{1}{2}}}\Theta V_a= -{\textstyle{\frac{1}{2}}}
\Theta\tilde{{\rm D}}_a\Phi\,,  \label{lin-Va-prop}
\end{equation}
where $\Phi=2\phi/a^{1/2}\Theta$ is the rescaled peculiar
gravitational potential. The above clearly manifests that
\begin{equation}
\dot{V}_a=0 \hspace{5mm}\Leftrightarrow\hspace{5mm}
V_a=-\tilde{{\rm D}}_a\Phi\,,  \label{free-motion}
\end{equation}
namely that to first order the rescaled peculiar velocity is
simply the gradient of the rescaled peculiar gravitational
potential. Furthermore, in the absence of background rotation,
result (\ref{free-motion}) also guarantees that
\begin{equation}
\tilde{{\rm D}}_{[b}V_{a]}= -\tilde{{\rm D}}_{[b}\tilde{{\rm
D}}_{a]}\Phi=0\,,  \label{irrot-motion}
\end{equation}
to leading order. Therefore, one can always treat the linear
peculiar motion of the dust component as acceleration-free and
irrotational.

\section{The second order peculiar motion of the matter}
\subsection{The original kinematical equations}
The linear equations of the previous section are adequate in the
early stages of gravitational aggregation when the perturbed
quantities are still relatively small. As the perturbations grow
stronger, however, the linear approximation brakes down and one
has to incorporate nonlinear effects. During the transition from
the linear to the fully nonlinear collapse, a period which one
might call ``the mildly nonlinear era'', one can adequately
monitor the perturbed variables by employing the second-order
equations instead of the fully nonlinear ones. In what follows, we
will employ these second order equations to study the growth of
velocity perturbations during the early nonlinear stages, as the
inhomogeneity decouples from the background expansion, and starts
to ``turn around'' and collapse under its own gravity.

Maintaining the Einstein-de Sitter background of the previous
section, we find that the second-order peculiar motion of the
pressure-free matter is monitored by the following set of
equations\footnote{We employ the linearization scheme used
in~\cite{MGE}. In brief, terms that vanish in the background, such
as the peculiar velocity or the electric Weyl tensor for example,
have perturbative order ${\cal O}(\epsilon)$, where $\epsilon$ is
the smallness parameter. To incorporate second order effects, we
include terms of order ${\cal O}(\epsilon^2,\,\epsilon v,\,v^2)$
and neglect those of order ${\cal O}(\epsilon^3,\,\epsilon
v^2,\,v^3)$ and higher. For more details the reader is referred
to~\cite{MGE}.}
\begin{eqnarray}
\dot{v}_{\langle a\rangle}&=&-A_{\langle a\rangle}-
{\textstyle{\frac{1}{3}}}\Theta v_{\langle a\rangle}\,,
\label{2va-prop}\\
\dot{\hat{\Theta}}&=&-{\textstyle{\frac{1}{3}}}\hat{\Theta}^2-
2\hat{\sigma}^2+ 2\hat{\omega}^2+ \tilde{{\rm D}}_a\dot{v}^a+
{\textstyle{\frac{1}{2}}}\mu v^2\,, \label{2p-Ray}\\
\dot{\hat{\sigma}}_{ab}&=&
-{\textstyle{\frac{2}{3}}}\hat{\Theta}\hat{\sigma}_{ab}-
\hat{\sigma}_{c\langle a}\hat{\sigma}_{b\rangle}{}^c-
\hat{\omega}_{c\langle a}\hat{\omega}_{b\rangle}{}^c+ \tilde{{\rm
D}}_{\langle a}\dot{v}_{b\rangle}\,,  \label{2p-shear-prop}\\
\dot{\hat{\omega}}_{ab}&=&
-{\textstyle{\frac{2}{3}}}\hat{\Theta}\hat{\omega}_{ab}+
2\hat{\sigma}_{c[a}\hat{\omega}_{b]}{}^c- \tilde{{\rm
D}}_{[a}\dot{v}_{b]}\,.  \label{2p-vort-prop}
\end{eqnarray}
These are obtained from Eqs.~(\ref{va-prop3})-(\ref{p-vort-prop3})
on using the linear transformation laws (\ref{lin-kin}) between
the kinematical quantities of the quasi-Newtonian and the matter
frame. We should also emphasize that the system
(\ref{2va-prop})-(\ref{2p-vort-prop}) applies to subhorizon scales
only, where the peculiar motion dominates over the background
expansion (i.e.~for $\Theta\ll\hat{\Theta}$).

Note that, to second perturbative order, the peculiar shear
evolves unaffected by tidal forces, since the electric and
magnetic Weyl tensors no longer contribute to
Eq.~(\ref{2p-shear-prop}). The absence of a shear-tide coupling is
a direct consequence of the two-frame approach we have employed.
It means that the evolution of the fluid element proceeds
unaffected by the long range gravitational field. This in turn has
crucial implications for the asymptotic final shape of the
collapsing overdensity. As we shall see later, the absence of the
electric tidal field means that generic collapse does not evolve
towards the spindle-like singularity that is usually associated
with the silent universes models~\cite{BJ,BMP1}.

\subsection{The rescaled equations}
In Sec.~3.3 we showed that, to linear order the peculiar motion of
the matter is also described by a rescaled velocity field which is
both acceleration-free and irrotational. The introduction of the
rescaled peculiar velocity $V_a=a^{-1/2}v_a$, leads immediately to
the rescaling of the irreducible kinematical variables of the
peculiar motion. In particular, given that $\tilde{{\rm
D}}_bv_a=a^{1/2}\tilde{{\rm D}}_bV_a$, we have
\begin{equation}
\hat{\Theta}=a^{1/2}\bar{\Theta}\,,\hspace{1cm}\hat{\sigma}_{ab}=a^{1/2}
\bar{\sigma}_{ab}\hspace{5mm}{\rm and}\hspace{5mm}
\hat{\omega}_{ab}=a^{1/2}\bar{\omega}_{ab}\,,  \label{p-kin2}
\end{equation}
where
\begin{equation}
\bar{\Theta}=\tilde{{\rm D}}_aV^a\,, \hspace{1cm}
\bar{\sigma}_{ab}=\tilde{ {\rm D}}_{\langle b}V_{a\rangle}
\hspace{5mm} {\rm and} \hspace{5mm} \bar{\omega}_{ab}=\tilde{{\rm
D}}_{[b}V_{a]}  \label{bar-kin}
\end{equation}
are respectively the rescaled peculiar contraction scalar, the
recsaled peculiar shear and the rescaled peculiar vorticity.

On using expressions (\ref{p-kin2}) one can transform
Eqs.~(\ref{2va-prop})-(\ref{2p-vort-prop}) into the following
system
\begin{eqnarray}
\dot{V}_{\langle a\rangle}&=&-a^{-1/2}A_{\langle a\rangle}-
{\textstyle{\frac{1}{2}}}\Theta V_{\langle a\rangle}\,,  \label{2Va-prop}\\
\dot{\bar{\Theta}}&=&-{\textstyle{\frac{1}{3}}}a^{1/2}\bar{\Theta}^2-
2a^{1/2}\bar{\sigma}^2+ 2a^{1/2}\bar{\omega}^2+ \tilde{{\rm D}}_a\dot{V}^a+
{\textstyle{\frac{1}{2}}}a^{1/2}\mu V^2\,,  \label{2pV-Ray}\\
\dot{\bar{\sigma}}_{ab}&=&-{\textstyle{\frac{2}{3}}}a^{1/2}
\bar{\Theta}\bar{\sigma}_{ab}- a^{1/2}\bar{\sigma}_{c\langle
a}\bar{\sigma}_{b\rangle}{}^c- a^{1/2}\bar{\omega}_{c\langle
a}\bar{\omega}_{b\rangle}{}^c+ \tilde{{\rm D}}_{\langle
a}\dot{V}_{b\rangle}\,,  \label{2pV-shear-prop}\\
\dot{\bar{\omega}}_{ab}&=&
-{\textstyle{\frac{2}{3}}}a^{1/2}\bar{\Theta}\bar{\omega}_{ab}+
2a^{1/2}\bar{\sigma}_{c[a}\bar{\omega}_{b]}{}^c- \tilde{{\rm
D}}_{[a}\dot{V}_{b]}\,,  \label{2pV-vort-prop}
\end{eqnarray}
which governs the motion of the matter, as described by the rescaled
peculiar velocity field $V_a$, to second order.

\section{The Zeldovich approximation}
\subsection{Propagation equations and constraints}
The Zeldovich approximation addresses the mildly nonlinear
collapse of protogalactic clouds, on scales well within the Hubble
radius, as they decouple from the background expansion and turn
around. The approximation builds upon the exact linear result of
acceleration-free and irrotational motion of the dust component
(see Sec.~3.3), which is extrapolated into the nonlinear regime.
This considerably simplifies the equations and allows one to
obtain analytical solutions. Here, we will describe the mildly
nonlinear regime through the set of the rescaled second order
equations (\ref{2Va-prop})-(\ref{2pV-vort-prop}). When the
Zeldovich ansatz $\dot{V}_a=0=\bar{\omega}_{ab}$ is applied, the
motion of the collapsing pressure-free matter is governed by
Eqs.~(\ref{2pV-Ray}) and (\ref{2pV-shear-prop}) only, which for
constant velocity and in the absence of vorticity reduce even
further to
\begin{eqnarray}
\dot{\bar{\Theta}}&=&-{\textstyle{\frac{1}{3}}}a^{1/2}\bar{\Theta}^2-
2a^{1/2}\bar{\sigma}^2+ {\textstyle{\frac{1}{2}}}a^{1/2}\mu V^2\,,
\label{Z-Ray} \\
\dot{\bar{\sigma}}_{ab}&=&
-{\textstyle{\frac{2}{3}}}a^{1/2}\bar{\Theta}\bar{\sigma}_{ab}-
a^{1/2}\bar{\sigma}_{c\langle a}\bar{\sigma}_{b\rangle}{}^c\,.
\label{Z-shear-prop}
\end{eqnarray}
At the same time, $\dot{V}_a=0=\bar{\omega}_{ab}$ means that
Eq.~(\ref{2Va-prop}) transforms into the constraint
\begin{equation}
A_a=-{\textstyle{\frac{1}{2}}}a^{1/2}\Theta V_a\,,
\label{Z-constraint}
\end{equation}
and that Eq.~(\ref{2pV-vort-prop}) becomes an identity. The latter
ensures that setting the vorticity to zero at second order is a
self-consistent assumption. This is not the case, however, with
constraint (\ref{Z-constraint}) whose consistency in time is not
straightforward. Nevertheless, constraint (\ref{Z-constraint})
does not pose any real problem for the consistency of the
Zeldovich ansatz at second order. Indeed, by propagating
(\ref{Z-constraint}) in time one obtains the following evolution
equation for the 4-acceleration vector
\begin{equation}
\dot{A}_a=-{\textstyle{\frac{1}{6}}}\Theta A_a+
{\textstyle{\frac{1}{4}}}a^{1/2}\mu V_a\,.  \label{Zconsistency}
\end{equation}
Since the zero-vorticity constraint is automatically satisfied at
second order, the above propagation equation is effectively the
consistency condition of the Zeldovich approximation to this
perturbative order.

We now turn our attention to the matter term at the end of
Eq.~(\ref{Z-Ray}), which conveys the effects of spacetime
curvature (see Sec.~2.3). Given that $V_a$ is constant and that
$\mu\propto a^{-3}$, the impact of the background matter upon
$\bar{\Theta}$ decays away. In other words, as the collapse is
progressively dominated by the kinematics and the role of gravity
becomes negligible. The situation is closely analogous to that
observed in studies of silent universe models
(e.g.~see~\cite{BMP1}), or in Bianchi~I cosmologies, where the
latter become shear dominated at early times
(e.g.~see~\cite{EvE}).

One should note that the effect of the approximation has been to
decouple the evolution along neighboring world lines from each
other, similar to what happens in silent universe models, where
(in an inhomogeneous situation) one can integrate the evolution
equations separately along neighboring worldlines. In other words,
the system of partial differential equations has been reduced to
an effective system of ordinary differential equations.

\subsection{The shear eigenframe}
To proceed further we introduce the new ``time'' variable $\tau $,
constructed so that $\dot{\tau}=-a^{1/2}\bar{\Theta}$, where the
minus sign compensates for the fact that we are dealing with a
collapsing region (i.e.~$\bar{\Theta}<0$). This choice guarantees
that $\dot{\tau}>0$ always and that $\tau\rightarrow\infty$ as we
approach the singularity and $\bar{\Theta}\rightarrow-\infty$.
Then, Eqs.~(\ref{Z-Ray}), without matter, and (\ref{Z-shear-prop})
transform into the following set
\begin{eqnarray}
\bar{\Theta}'&=&{\textstyle{1\over3}}\bar{\Theta}+
2\bar{\Theta}^{-1}\bar{\sigma}^2\,,  \label{Z-Ray2}\\
\bar{\sigma}'_{ab}&=&{\textstyle{2\over3}}\bar{\sigma}_{ab}+
\bar{\Theta}^{-1}\bar{\sigma}_{c\langle
a}\bar{\sigma}_{b\rangle}{}^c\,,  \label{Z-shear-prop2}
\end{eqnarray}
where from now on a prime will indicate differentiation with
respect to $\tau$. Assuming the shear eigenframe we have
$\bar{\sigma}_{ab}=(\bar{\sigma}_{11},\, \bar{\sigma}_{22},\,
\bar{\sigma}_{33})$, where
$\bar{\sigma}_{33}=-(\bar{\sigma}_{11}+\bar{\sigma}_{22})$ due to
the trace-free nature of the shear. Then, if
$\bar{\sigma}_1=\bar{\sigma}_{11},\,
\bar{\sigma}_2=\bar{\sigma}_{22}$ and
$\bar{\sigma}_3=\bar{\sigma}_{33}$ determine the three shear
eigen-directions, the second order system (\ref{Z-Ray2}),
(\ref{Z-shear-prop2}) reads
\begin{eqnarray}
\bar{\Theta}'&=&{\textstyle{1\over3}}\bar{\Theta}+
2\bar{\Theta}^{-1}(\bar{\sigma}_1^2+\bar{\sigma}_2^2
+\bar{\sigma}_1\bar{\sigma}_2)\,,
\label{Z-Ray3}\\
\bar{\sigma _1}'&=&{\textstyle{2\over3}}\bar{\sigma}_1+
{\textstyle{1\over3}}\bar{\Theta}^{-1}\bar{\sigma}_1^2-
{\textstyle{2\over3}}\bar{\Theta}^{-1}(\bar{\sigma}_1
+\bar{\sigma}_2)\bar{\sigma}_2\,,
\label{Z-shear1-prop3}\\
\bar{\sigma_2}'&=&{\textstyle{2\over3}}\bar{\sigma}_2+
{\textstyle{1\over3}}{\bar{\Theta}}^{-1}\bar{\sigma}_2^2-
{\textstyle{2\over3}}\bar{\Theta}^{-1}(\bar{\sigma}_1
+\bar{\sigma}_2)\bar{\sigma}_1\,, \label{Z-shear2-prop3}
\end{eqnarray}
where the behavior of $\bar{\sigma}_3$ is determined from that of
$\bar{\sigma}_1$, $\bar{\sigma}_2$ entirely. The above
second-order equations govern the small-scale evolution of the
pressure-free matter, as the latter decouples from the background
expansion and starts to turn around, and provide a fully covariant
formulation of the Zeldovich approximation.

The adoption of the shear eigenframe means that the three shear
eigen-directions acquire special importance. Consequently, it is
convenient to introduce the quantities
\begin{equation}
\bar{\Theta}_i=\bar{\sigma}_i+
{\textstyle{1\over3}}\bar{\Theta}\,, \label{bThetai}
\end{equation}
with $i=1,2,3$, which provide the contraction rates in each shear
eigen-direction. They will also be used to define a scaling length
along each direction (see Sec.~5.4).

\subsection{The attractors}
The question is whether or not the relativistic analysis also
predicts that one-dimensional pancakes are the final
configurations of any generic collapsing overdensity. Given the
qualitative nature of the question one can employ a dynamical
system approach to provide the answer. We begin by defining the
following dimensionless dynamical variables (see also~\cite{Br,W})
\begin{equation}
\Sigma_{+}=\frac{3(\bar{\sigma}_{1}
+\bar{\sigma}_{2})}{2\bar{\Theta}} \hspace{10mm} {\rm and}
\hspace{10mm} \Sigma_{-}=\frac{\sqrt{3}(\bar{\sigma}_{1}
-\bar{\sigma}_{2})}{2\bar{\Theta}}\,, \label{Sigmas}
\end{equation}
which in turn imply the auxiliary relations
\begin{equation}
\bar{\sigma}_{1}={\textstyle{1\over3}}\bar{\Theta}(\Sigma_{+}
+\sqrt{3}\Sigma_{-})\,, \hspace{5mm}
\bar{\sigma}_{2}={\textstyle{1\over3}}\bar{\Theta}(\Sigma_{+}
-\sqrt{3}\Sigma_{-}) \hspace{5mm} {\rm and} \hspace{5mm}
\bar{\sigma}_{3}=-{\textstyle{2\over3}}\bar{\Theta}\Sigma_{+}\,.
\label{aux}
\end{equation}
Clearly, $\Sigma _{+}$ and $\Sigma _{-}$ measure the kinematical
anisotropy of the collapse. If both are zero the collapse is
spherically symmetric. When only $\Sigma_{-}=0$ we have the
degenerate case case of two equal shear eigenvalues. On
introducing $\Sigma_{\pm}$ Eq.~(\ref{Z-Ray3}) transforms into
\begin{equation}
\bar{\Theta}'={\textstyle{1\over3}}\bar{\Theta}+
{\textstyle{2\over3}}\bar{\Theta}\left(\Sigma_{+}^2+
\Sigma_{-}^2\right)\,, \label{Z-Ray4}
\end{equation}
while Eqs.~(\ref{Z-shear1-prop3}) and (\ref{Z-shear2-prop3})
read~\cite{Br}
\begin{eqnarray}
\Sigma'_{+}&=&{\textstyle{1\over3}}
\left[1-\Sigma_{+}-2(\Sigma_{+}^2+\Sigma_{-}^2)\right]\Sigma_{+}+
{\textstyle{1\over3}}\Sigma_{-}^2\,,  \label{Z-Sigma1-prop}\\
\Sigma'_{-}&=&{\textstyle{1\over3}}\left[1+2\Sigma_{+}-2(\Sigma_{+}^2
+\Sigma_{-}^2)\right]\Sigma_{-}\,. \label{Z-Sigma2-prop}
\end{eqnarray}
Accordingly, the evolution of $\Sigma _{+}$ and $\Sigma _{-}$ has
decoupled from that of $\bar{\Theta}$ and the shape evolution of
the overdensity is now monitored by the sub-system
(\ref{Z-Sigma1-prop}), (\ref{Z-Sigma2-prop}). Technically
speaking, the problem has now been reduced to the study of the
planar dynamical system (\ref {Z-Sigma1-prop}),
(\ref{Z-Sigma2-prop}) depicted in Fig.~1. Physically, the
dimensional reduction means that the shape of the collapsing dust
cloud does not depend on the time scale of the collapse. Note that
our equations are invariant under the cyclic symmetry

\begin{figure}[tbp]
\centering \vspace{8cm} \includegraphics{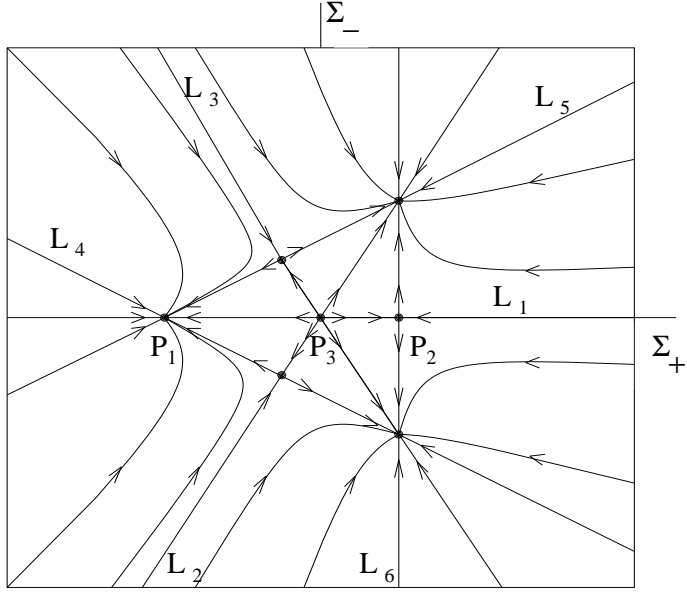} \caption{Phase plane $\Sigma_{+}\equiv X$,
$\Sigma_{-}\equiv Y$. The lines forming the central triangle
correspond to $\bar{\sigma}_i=-\bar{\Theta}/3$ (with $i=1,2,3$).
The three pancakes are located at the intersections of these
lines. The lines bisecting the vertices correspond to two equal
shear eigenvalues, (e.g.~$\Sigma_{-}=0$ corresponds to
$\bar{\sigma}_1=\bar{\sigma}_2$)~\cite{Br}.} \label{fig:1}
\end{figure}

\begin{equation}
\bar{\sigma}_{1} \;\; \rightarrow \;\; \bar{\sigma}_{2} \;\;
\rightarrow \;\; \bar{\sigma}_{3} \;\; \rightarrow \;\;
\bar{\sigma}_{1}\,,
\end{equation}
which is equivalent to the chain
\begin{eqnarray}
\Sigma_{+} \;\; \rightarrow \;\;
-{\textstyle{1\over2}}\Sigma_{+}-{\textstyle{\sqrt{3}\over2}}\Sigma_{-}
\;\; \rightarrow \;\; -{\textstyle{1\over2}}\Sigma
_{+}+{\textstyle{\sqrt{3}\over2}}\Sigma_{-} \;\; \rightarrow \;\;
\Sigma _{+}\,,\\ \Sigma_{-} \;\; \rightarrow \;\;
+{\textstyle{\sqrt{3}\over2}}\Sigma_{+}-{\textstyle{1\over2}}\Sigma_{-}
\;\; \rightarrow \;\;
-{\textstyle{\sqrt{3}\over2}}\Sigma_{+}-{\textstyle{1\over2}}\Sigma_{-}
\;\; \rightarrow \;\; \Sigma_{-}\,.
\end{eqnarray}

The two sets of lines in Fig.~1, those that form the central
triangle and those that bisect the vertex angles, are each
invariant submanifolds (mapped into themselves by the flow). The
{\it bisecting lines} represent the rotationally symmetric
solutions with trajectories
\begin{eqnarray}
L1: &\;\Sigma_{-}=0 &\Leftrightarrow \;\;\;
\bar{\sigma}_{1}=\bar{\sigma}_{2}\,,\\ L2: &\;\Sigma
_{-}=\sqrt{3}\Sigma_{+} &\Leftrightarrow \;\;\;
\bar{\sigma}_{2}=\bar{\sigma}_{3}\,,\\ L3:
&\;\Sigma_{-}=-\sqrt{3}\Sigma_{+} &\Leftrightarrow\;\;\;
\bar{\sigma}_{1}=\bar{\sigma}_{3}\,.
\end{eqnarray}
They are mapped into themselves by the symmetries noted above, so
they are equivalent to each other. Concentrating on $L1$, we
notice that it is involutive because $\Sigma_{-}=0$ immediately
implies $\Sigma'_{-}=0$ (see Eq.~(\ref{Z-Sigma2-prop})). The
equation along this trajectory is
\begin{equation}
\Sigma_{+}'={\textstyle{1\over3}}
\left[1-\Sigma_{+}-2\Sigma_{+}^2\right]\Sigma_{+}\,, \hspace{5mm}
\Sigma_{-}={0}
\end{equation}
representing a family of rotationally symmetric solutions. It has
three fixed (stationary) points, which are each self-similar
solutions of the system (\ref {Z-Sigma1-prop}),
(\ref{Z-Sigma2-prop}). In particular we have\vspace{5mm}

P1~($\Sigma_{-}=0,\Sigma_{+}=-1$): In this case
$\bar{\sigma}_1=\bar{\sigma}_2=-\bar{\Theta}/3$ and
$\bar{\sigma}_3=2\bar{\Theta}/3$, which implies that
$\bar{\Theta}_1=\bar{\Theta}_2=0$ and
$\bar{\Theta}_3=\bar{\Theta}$. This point corresponds to a
rotationally symmetric pancake collapse. It occurs at a triangle
vertex, where two of the triangle generating lines and a bisecting
line all intersect, and is an attractor in this orbit.\vspace{5mm}

P2~($\Sigma_{-}=0,\Sigma_{+}=\frac 12$): Here
$\bar{\sigma}_1=\bar{\sigma}_2=\bar{\Theta}/6$ and
$\bar{\sigma}_3=-\bar{\Theta}/3$ meaning that
$\bar{\Theta}_1=\bar{\Theta}_2=\bar{\Theta}/2$ and $
\bar{\Theta}_3=0$. This point corresponds to a spindle or
filamentary rotationally symmetric collapse. It occurs where a
bisecting line meets a triangle generating line, and is an
attractor in the orbit.\vspace{5mm}

P3~($\Sigma_{-}=0,\Sigma_{+}=0$): At this point
$\bar{\sigma}_1=\bar{\sigma}_2=\bar{\sigma}_3=0$ and
$\bar{\Theta}_1=\bar{\Theta}_2=\bar{\Theta}_3=\bar{\Theta}/3$,
which corresponds to an isotropic collapse. It occurs at the
centre where the three bisecting lines meet, and is a source in
the orbit. Note that the same set of fixed points also occur on
each of the other two bisecting lines $L2$ and $L3$.\vspace{5mm}

The {\it triangle generating lines} represent the (generally anisotropic)\
2-dimensional collapse solutions with trajectories
\begin{eqnarray}
L4: &\;\bar{\Theta}_{1}=\bar{\sigma}_{1}+
{\frac{1}{3}}\bar{\Theta}=0 &\Leftrightarrow \;\;\;
\Sigma_{-}=-{\textstyle{1\over\sqrt{3}}}\left(\Sigma_{+}+1\right)\,,\\
L5: &\;\bar{\Theta}_{2}=\bar{\sigma}_{2}+
{\frac{1}{3}}\bar{\Theta}=0 &\Leftrightarrow \;\;\;
\Sigma_{-}=+{\textstyle{1\over\sqrt{3}}}\left(\Sigma_{+}+1\right)\,,\\
L6: &\;\bar{\Theta}_{3}=\bar{\sigma}_{3}+
{\frac{1}{3}}\bar{\Theta}=0 &\Leftrightarrow \;\;\;
\Sigma_{+}={\textstyle{1\over2}}\,.
\end{eqnarray}
They are mapped into themselves by the symmetries noted above, so
they are equivalent to each other. Let us concentrate on $L6$,
which is involutive since $\Sigma_{+}=1/2$ leads to
$\Sigma'_{+}=0$ (see Eq.~(\ref{Z-Sigma1-prop})). The equation
along this trajectory is
\begin{equation}
\Sigma_{+}={\textstyle{1\over2}}\,, \hspace{5mm}
\Sigma'_{-}={\textstyle{2\over3}}
\left[{\textstyle{3\over4}}-\Sigma_{-}^{2}\right]\Sigma_{-}\,,
\end{equation}
representing a family of exact 2-dimensional collapse solutions.
The latter are rotationally symmetric only where they intersect
the bisecting lines. This orbit has three fixed points. The point
($\Sigma_{-}=0,$ $\Sigma_{+}=1/2$), where it intersects a
bisecting line, namely the LRS spindle P2 noted above. Also, the
two points ($\Sigma_{-}=\sqrt{3}/2$, $\Sigma_{+}=1/2$) and
($\Sigma_{-}=-\sqrt{3}/2$, $\Sigma_{+}=1/2)$, where L6 meets
another generating line and a bisecting line simultaneously. Each
of the latter two points is equivalent to the pancake collapse
node P1 noted above.

In summary, the vertices of the triangle are stationary points of
the system (\ref{Z-Sigma1-prop}), (\ref{Z-Sigma2-prop}) and are
attractors. Each is a one-dimensional pancake solution, with a
stationary state along two directions and collapse along the third
shear eigen-direction. Generic solutions are asymptotic to one of
these three one-dimensional pancakes (one for each
eigen-direction) and so are asymptotic to one-dimensional
pancakes. The bisecting lines intersect at the point that
corresponds to a shear-free spherically symmetric collapse, also a
stationary point. Where the bisecting lines intersect the
triangle, we have stationary points that represent exact
filamentary solutions. The pancakes are stable nodes, the
filaments are saddle points, and the spherically symmetric
collapse is an unstable node. This proves that, once collapse sets
in, the pancakes are the natural attractors for a generic
overdensity.

\subsection{The stationary points}
In cosmology we use the average expansion scalar to define the
cosmological scale factor. Similarly, one can utilize the collapse
rate ($\hat{\Theta}$) of an inhomogeneity to define an average
local scaling length ($\ell$) as well as directional scaling
lengths ($\ell_i$, $i=1,2,3$) along the three shear
eigen-directions (e.g.~see~\cite{BMP1}). In particular we have
\begin{equation}
\frac{\dot{\ell}}{\ell}={\textstyle{1\over3}}\hat{\Theta}
\hspace{10mm} {\rm and} \hspace{10mm}
\frac{\dot{\ell}_i}{\ell_i}=\hat{\sigma}_i+
{\textstyle{1\over3}}\hat{\Theta}=\hat{\Theta}_i\,.  \label{ells}
\end{equation}
Thus, $\ell$ is the geometric mean of the three directional scale
factors $\ell_i$, while $\hat{\Theta}$ is the average of their
expansion rates. Following the standard terminology of~\cite{Mac},
pancakes occur when two of the $\ell_i$s approach finite vales
while the other goes to zero, spindle-like singularities
correspond to two of the $\ell_i$s going to zero and the third to
a constant, and finally a point-like singularity occurs when all
three of the $\ell_i$s tend to zero.\vspace{5mm}

As we have already pointed out, stationary points correspond to
finite values of the dimensionless dynamical variables
$\Sigma_{\pm}$ introduced in Sec.~5.3. Concentrating on the three
stationary points along the $L1$ trajectory, we can calculate the
associated collapse timescales. In particular, following closely
the analysis given in~\cite{BMP1} we have:\vspace{5mm}

P1~($\Sigma _{-}=0,\Sigma _{+}=-1$): This case corresponds to
pancake collapse along the third shear eigen-direction,
Eq.~(\ref{Z-Ray4}) implies that $\bar{\Theta}'=\bar{\Theta}$.
Here, the evolution of the contraction scalar $\hat{\Theta}$ is
given by
\begin{equation}
\hat{\Theta}=\frac{4\hat{\Theta}_0t^{1/3}}
{4t_0^{1/3}+3\hat{\Theta}_0(t^{4/3}-t_0^{4/3})}\,. \label{pTheta}
\end{equation}
where $t$ is the proper time along the matter frame and the zero
suffix indicates the beginning of the collapse. Note that we have
used the transformation law (\ref{p-kin2}a) and assumed a
dust-dominated universe (i.e.~$a\propto t^{2/3}$). Then, by
integrating Eqs.~(\ref{ells}) we find that the average scale
factor and the scaling length along the direction of the collapse
evolve as
\begin{equation}
\ell=\ell_0\left[1+\alpha_1(t^{4/3}-t_0^{4/3})\right]^{1/3}
\hspace{5mm} {\rm and} \hspace{5mm}
\ell_3=(\ell_3)_0\left[1+\alpha_1(t^{4/3}-t_0^{4/3})\right]\,,
\label{pells}
\end{equation}
respectively. Note that $\ell_1$, $\ell_2=$ constant and
$\alpha_1=3\hat{\Theta}_0/4t^{1/3}<0$, given that
$\hat{\Theta}_0<0$ for collapse. Thus, the pancake singularity
occurs at
\begin{equation}
t_*=t_0^{1/4}\left(t_0
-{\textstyle{4\over3}}\hat{\Theta}_0^{-1}\right)^{3/4}\,.
\label{pt*}
\end{equation}\vspace{5mm}

P2~($\Sigma _{-}=0,\Sigma _{+}=\frac 12$): For filamentary
collapse along the first two shear eigen-directions
Eq.~(\ref{Z-Ray4}) gives $\bar{\Theta}'=\bar{\Theta}/2$. This
leads to the following evolution law
\begin{equation}
\hat{\Theta}=\frac{8\hat{\Theta}_0t^{1/3}}
{8t_0^{1/3}+3\hat{\Theta}_0(t^{4/3}-t_0^{4/3})}\,, \label{fTheta}
\end{equation}
for the average contraction scalar. In addition, we find that the
average and the two directional scaling lengths evolve as
\begin{equation}
\ell=\ell_0\left[1+\alpha_2(t^{4/3}-t_0^{4/3})\right]^{2/3}\,,
\label{fell}
\end{equation}
\begin{equation}
\ell_1=(\ell_1)_0\left[1+\alpha_2(t^{4/3}-t_0^{4/3})\right]
\hspace{5mm} {\rm and} \hspace{5mm}
\ell_2=(\ell_2)_0\left[1+\alpha_2(t^{4/3}-t_0^{4/3})\right]\,,
\label{fells}
\end{equation}
respectively, while $\ell_3=$ constant
($\alpha_2=3\hat{\Theta}_0/8t^{1/3}$). According to the above, the
spindle-like singularity takes place at
\begin{equation}
t_*=t_0^{1/4}\left(t_0
-{\textstyle{8\over3}}\hat{\Theta}_0^{-1}\right)^{3/4}\,.
\label{ft*}
\end{equation}\vspace{5mm}

P3~($\Sigma _{-}=0,\Sigma _{+}=0$): Finally, for isotropic
collapse $\bar{\Theta}'=\bar{\Theta}/3$ (see Eq.~(\ref{Z-Ray4}))
which in turn implies
\begin{equation}
\hat{\Theta}=\frac{4\hat{\Theta}_0t^{1/3}}
{84t_0^{1/3}+\hat{\Theta}_0(t^{4/3}-t_0^{4/3})} \hspace{5mm} {\rm
and} \hspace{5mm}
\ell=\ell_0\left[1+\alpha_3(t^{4/3}-t_0^{4/3})\right]
\label{iTheta-ell}
\end{equation}
for the contraction scalar and the scaling length respectively
($\alpha_3=\hat{\Theta}_0/4t^{1/3}$). Here, the point singularity
occurs at
\begin{equation}
t_*=t_0^{1/4}\left(t_0-4\hat{\Theta}_0^{-1}\right)^{3/4}\,.
\label{it*}
\end{equation}

Clearly, the contracting behavior at each stationary point is
monotonic and fixed by the initial conditions. Given that
$\hat{\Theta}_0<0$, there is always a singularity in the future.
According to Eqs.~(\ref{pt*}), (\ref{ft*}) and (\ref{it*}), for
the same set of initial conditions, the pancake singularity occurs
first and the point singularity last. This result, which is due to
the increase in the contraction rate of fluid congruence in the
presence of shear, is in agreement with the qualitative ``collapse
theorem'' given in~\cite{BJ}.

\section{Conclusions}
A key issue in contemporary theoretical cosmology is understanding
the physics of gravitational collapse and the mechanisms that have
given rise to the observed large-scale structure of the universe.
Linear perturbations, about a Friedmann-Robertson-Walker model,
are fairly straightforward to follow. When certain simplifying
symmetries are imposed, the nonlinear collapse can also be treated
analytically. The spherical top-hat model, in Newtonian theory,
and the Tolman-Bondi solution, in General Relativity, are such
examples. When it comes to the more realistic non-spherical
collapse, however, the Zeldovich approximation has been the most
influential and celebrated paradigm. It has also inspired a number
of variations, all of which try to address the complexities of
structure evolution beyond ``caustic'' formation. Despite the
number of these different treatments, however, a fully covariant
approach to the Zeldovich approximation has been missing. This is
the issue that the present paper has tried to address.

We have pursued a relativistic treatment of nonlinear peculiar
velocities by adopting a two-frame approach which enables us to
provide a truly Lagrangian formalism. Assuming matter with
non-relativistic peculiar motion relative to a quasi-Newtonian
frame, we have derived the fully nonlinear equations and applied
them to the case of pressure-free dust. The inherent transparency
of the covariant formalism means that all the terms in our
equations have a clear physical and geometrical interpretation.
For a given background, it is also straightforward to identify the
perturbative order of all the effects that influence the peculiar
motion of the matter. Assuming an Einstein-de Sitter background
and allowing for up to second order perturbative terms, we have
addressed the mildly nonlinear collapse and in the process
provided a fully covariant formulation of the Zeldovich
approximation. On introducing the Zeldovich ansatz of
acceleration-free and irrotational peculiar motion, our equations
have reduced to a planar dynamical system, which has
one-dimensional pancakes as its natural attractors. Thus, just
like the Newtonian treatment, the relativistic approach also
predicts that, as long as we consider the mildly nonlinear stage,
pancakes are the final end-states of any generic collapsing
overdensity. In addition, by looking at the collapse timescales
associated with the stationary points, we have found that pancake
collapse takes place first and point-like singularities occur
last.

Our results for the fate of generic collapsing overdensities are
in disagreement with studies of silent-universe dynamics, which
argue for spindle-like rather than pancake singularities
(see~\cite{BJ,BMP1}). The reason for this difference lies in the
role of the tidal field within the adopted cosmological model.
Silent universes allow for non-zero electric Weyl tensor, but set
its magnetic counterpart to zero. In the presence of this
``truncated'' tidal field the fluid element evolves towards a
Kasner-type singularity, where pancakes are a set of measure zero.
Our two-frame analysis of a perturbed Einstein-de Sitter model, on
the other hand, shows that the influence of the long-range
gravitational field is negligible at second order. Technically
speaking, the covariant equations governing the mildly nonlinear
collapse of a dust cloud have zero Weyl-curvature input,
effectively reducing to the Newtonian ones. As a result, pancakes
are reinstated as the natural attractors of a generic collapsing
overdensity, this time within the relativistic framework. This is
in agreement with numerical simulations which clearly favor
pancake formation over all other types of structures
(see~\cite{SMMPT}). Once again we point out that our result holds
at second perturbative order and refers to what is known as the
mildly nonlinear regime. Although the second order equations can
probe deep into the epoch of structure formation, they are
expected to break down close to the singularity. Clearly, if we
were to address the full picture, we need to incorporate the tide
effects. For an Einstein-de Sitter background, however, the
electric and of the magnetic Weyl terms are both of third
perturbative order (see Eq.~(\ref{p-shear-prop3})). In other
words, the fully nonlinear collapse evolves under the influence of
the full tidal field rather than the purely-electric Weyl field of
the silent models. In addition, at higher than the second
perturbative order, the magnetic Weyl tensor is also a source of
peculiar vorticity (see Eqs.~(\ref{p-vort-prop2}),
(\ref{p-vort-prop3})) and one can no longer consistently argue for
irrotational peculiar motion. Of course, as the collapse proceeds
beyond the mildly nonlinear stage, pressure gradients also become
important, which means that an acceleration-free motion is no
longer sustainable, and the whole idea of the Zeldovich
approximation is expected to break down.

Note that, although it is usually bypassed, vorticity is an
important issue, given the observed rotation of galaxies and
galaxy clusters. By construction, the Zeldovich approximation
cannot address this question. Newtonian modifications of the
standard approach to incorporate vorticity have already been
suggested in the past (see~\cite{BS}). The covariant formalism
developed here can provide the basic framework for a relativistic,
mathematically rigorous and physically transparent treatment of
rotating dust during the nonlinear collapse.

\section{Acknowledgements}
We would like to thank John Barrow, Marco Bruni, Peter Dunsby,
Charles Hellaby and Henk van Elst for very helpful discussions and
comments. CGT also thanks Henk van Elst for his hospitality at
QMUL and Uli Kirchner for his help with the graphics. Figure 1 is
from reference~\cite{Br} and we wish to thank Marco Bruni for
giving us his permission to use it. CGT was supported by a
Sida/NRF grant.

\end{document}